%% file: CSCRYOSEM.tex
\documentclass{article}
\input{packages}
\begin{document}
\title{A Targeted Sampling Strategy for Compressive \\Cryo Focused Ion Beam Scanning Electron Microscopy}

\author{
	D.~Nicholls\textsuperscript{1},  
	J.~Wells\textsuperscript{2},
	A.~W.~Robinson\textsuperscript{1},
	A.~Moshtaghpour\textsuperscript{1, 3},
	M.~Kobylynska\textsuperscript{4},\\
	R.~A.~Fleck\textsuperscript{4},
	A.~I.~Kirkland\textsuperscript{3,5}, and N.~D.~Browning\textsuperscript{1,6,7}
	\footnote{
		DN, AWR, JW, NDB are funded by EPSRC. DN and JW are also funded by Sivananthan Laboratories. AM and AIK are funded by RFI.
		}
	\\
	\fontsize{9}{1}\selectfont \textsuperscript{1} Department of Mechanical, Materials and Aerospace Engineering, University of Liverpool, UK.\\
	\fontsize{9}{1}\selectfont \textsuperscript{2} Distributed Algorithms Centre for Doctoral Training, University of Liverpool, UK.\\
	\fontsize{9}{1}\selectfont \textsuperscript{3} Correlated Imaging group, Rosalind Franklin Institute, Harwell Science and Innovation Campus, Didcot, UK.\\
	\fontsize{9}{1}\selectfont \textsuperscript{4} Centre for Ultrastructural Imaging, King’s College London, London, UK.\\
	\fontsize{9}{1}\selectfont \textsuperscript{5} Department of Materials, University of Oxford, UK.\\
	\fontsize{9}{1}\selectfont \textsuperscript{6} Physical and Computational Science Directorate, Pacific Northwest National Laboratory, Richland, USA.\\
	\fontsize{9}{1}\selectfont \textsuperscript{7} Sivananthan Laboratories, 590 Territorial Drive, Bolingbrook, IL, USA.}\date{\empty}
	
%

\maketitle
\begin{abstract}
\noindent
Cryo Focused Ion-Beam Scanning Electron Microscopy (cryo FIB-SEM) enables three-dimensional and nanoscale imaging of biological specimens via a slice and view mechanism. The FIB-SEM experiments are, however, limited by a slow (typically, several hours) acquisition process and the high electron doses imposed on the beam sensitive specimen can cause damage. In this work, we present a compressive sensing variant of cryo FIB-SEM capable of reducing the operational electron dose and increasing speed. We propose two Targeted Sampling (TS) strategies that leverage the reconstructed image of the previous sample layer as a prior for designing the next subsampling mask. Our image recovery is based on a blind Bayesian dictionary learning approach, \ie Beta Process Factor Analysis (BPFA). This method is experimentally viable due to our ultra-fast GPU-based implementation of BPFA. Simulations on artificial compressive FIB-SEM measurements validate the success of proposed methods: the operational electron dose can be reduced by up to 20 times. These methods have large implications for the cryo FIB-SEM community, in which the imaging of beam sensitive biological materials without beam damage is crucial.

\end{abstract}
\vspace{2mm}
\section{Introduction}
\label{sec:intro}
Focused Ion-Beam Scanning Electron Microscopy (FIB-SEM) tomography is an instrument capable of producing a three-dimensional (3D) data cube of a material through SEM imaging of a series of successive FIB cross-sections (\ie slices): a FIB is used to remove a layer of material followed by an SEM sensing process to acquire an image \cite{giannuzzi2004introduction,kizilyaprak2014fib}. This process is then repeated for as many slices as desired. While this technique is very useful, it can only be applied to materials which fit certain criteria. For example, the material must be beam stable, otherwise the contact with the ion and/or electron beams causes the structures of the sample to change, leading to inaccurate analysis. In life sciences, these samples are commonly chemically fixed and embedded in resin to impart resistance to the beam \cite{fleck2019brief}.

\begin{figure}[t]
    \centering
    \scalebox{0.8}{\includegraphics{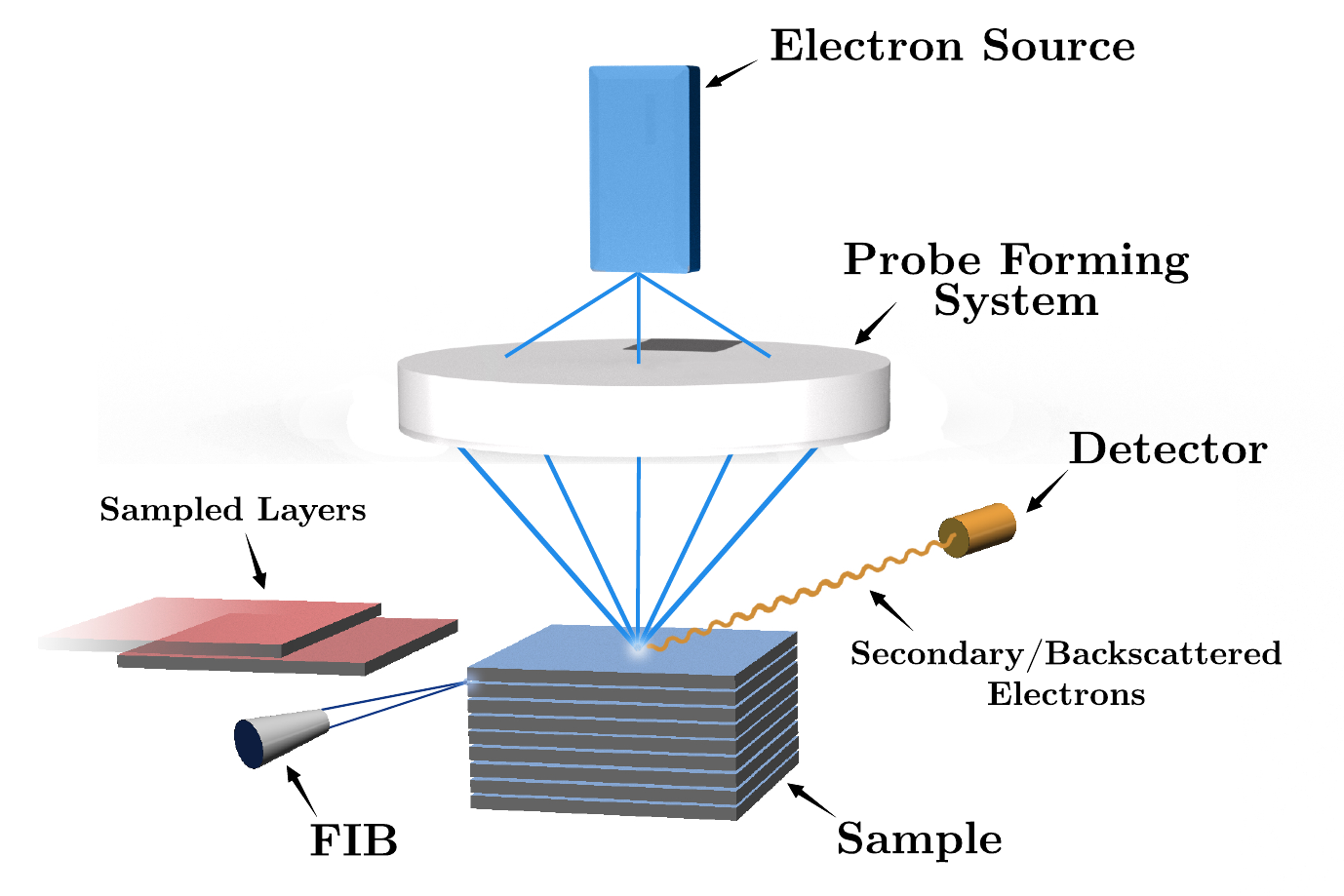}}
    \caption{Operating principles of Cryo FIB-SEM. A focused ion-beam is used to remove a layer of material. A scanning electron microscope is then used to obtain surface information from that newly revealed surface.}
    \label{fig:fib-sem-scheme}
\end{figure}

Fitting the FIB-SEM with cryo-technology allows the large volume ($\sim$100$\mu$m\textsuperscript{3}) analysis of biological specimens suspended in vitreous ice \cite{hayles2021introduction, schertel2013cryo}. This stabilises and perfectly preserves the structures of the specimen in their native state \cite{fleck2019brief, fleck2015low}. This also allows more accurate signal acquisition during experiment whilst also reducing the effects of beam induced damage during exposure critically enable the practical adoption of this powerful technique.

Alongside beam damage, experimental times, which are proportional to the number of cross-sections and images (hundreds of images are typically required), typically lie on the timescale of hours. This severely limits the data throughput rate of the FIB-SEM, more-so than any other factor.

The theory of Compressive Sensing (CS)~\cite{donoho2006compressed,candes2006robust} can aid with the aforementioned challenges. CS has been shown in SEM \cite{anderson2013sparse} and scanning transmission electron microscopy (STEM) to decrease beam damage and acquisition time whilst maintaining sufficient image quality \cite{binev2012compressed, stevens2018subsampled, donati2017compressed, nicholls2020minimising,  nicholls2021subsampled}.

In this paper we introduce a first compressive cryo FIB-SEM approach. Leveraging the sequential acquisition process of FIB-SEM, we develop a sampling strategy, referred to as \textit{targeted sampling}, based on CS theory. Each mask is generated from the previous image in a series, which takes advantage of the similarity between any two successive frames. The recovery method for each frame was performed using an ultra-fast GPU implementation of the Beta-Process Factor Analysis (BPFA). This novel implementation allows for real-time image recovery, which enables this method to be practically viable.  BPFA has been used in compressive STEM \cite{kovarik2016implementing, nicholls2022compressive}, with SEM imaging facing some of the same challenges as STEM imaging. 

This study aims to offer a solution for low-dose budget cryo FIB-SEM through compressive sensing methods with the intention of lowering induced beam damage and acquisition times.

\section{Methods: Compressive Cryo FIB-SEM}

\label{sec:cs-sem}
In this section we describe the proposed cryo FIB-SEM that collects compressed SEM measurements by subsampling the electron probe positions for each layer of the sample and reconstructs the thick sample image from those measurements by solving a Bayesian dictionary learning problem.

\subsection{Acquisition model}
Let $ \bs X \in \bb R^{N_1\times N_2\times N_3}$ be the discretised 3D representation of the sample with $\bs X_l \in \bb R^{N_1 \times N_2}$, for $l \in \{1, \cdots, N_3\}$, being a two-dimensional (2D) slice of $\bs X$ along the third dimension. In the following we assimilate $\bs X_l$ to its vectorised version $\bs x_l \coloneqq {\rm vec}(\bs X_l)\in \bb R^{\bar{N}}$, with $\bar{N} = N_1 N_2$. As explained in Sec.~\ref{sec:intro}, the FIB module removes a layer of the sample and allows the SEM to image the surface of the next layer. When operating in the full-acquisition mode, \eg raster scan, the electron probe scans over every location of the current sample surface, as shown in Fig.~\ref{fig:fib-sem-scheme}, and the resulting secondary/back-scattered electrons are collected by the detector. This simplified cryo FIB-SEM sensing model can be mathematically written as
\begin{equation}\label{eq:full-fib-sem}
    \bs y_l^{\rm full} \coloneqq \bs x_l + \bs n_l^{\rm full} \in \bb R^{\bar{N}},
\end{equation}
for every layer $l \in \{1,\cdots,N_3\}$, where $\bs n_l^{\rm full}$ is a noise. We here introduce a compressive cryo FIB-SEM that reduces the electron dose imposed on the sample. This is achieved by subsampling $M_l < \bar{N}$ probe positions (over the $l-$th sample surface) collected in a subsampling set $\Omega_l \subset \{1,\cdots,\bar{N}\}$ of size $|\Omega_l| = M_l$;  resulting in the following model
\begin{equation}\label{eq:cs-fib-sem}
    \bs y_l^{\rm cs} \coloneqq \bs P_{\Omega_l}\bs x_l + \bs n_l^{\rm cs} \in \bb R^{\bar{N}},
\end{equation}
for every layer $l \in \{1,\cdots,N_3\}$, where $\bs P_\Omega \in \{0,1\}^{\bar{N}\times \bar{N}}$ is a mask operator with $(\bs P_\Omega \bs x)_j = x_j$ if $j \in \Omega$, and  $(\bs P_\Omega \bs x)_j = 0$ otherwise. While the sensing model in \eqref{eq:cs-fib-sem} is general, in this paper we assume that the same number of probe positions are sampled for every layer, \ie $M_l = M$ for all $l$, and leave the study of the general model to a future paper.

Since the subsampling is performed in the image domain, a natural choice in subsampling strategy of probe positions is Uniform Density Sampling (UDS): for every layer, $M$ indices are selected (without replacement) with respect to (w.r.t) the uniform distribution. 

However, in this paper we adopt the \textit{Targeted Sampling (TS)} strategy introduced in \cite{robinson2022sim} for simulating electron microscopy images. For sampling layer $l \in \{1,\cdots,N_3\}$, given the number of measurements $M$ and the TS parameter $\rho \in [0,1]$, we select $M_{\rm t} \coloneqq \left \lfloor  \rho M\right \rfloor$ indices (\ie targeted part) in $\Omega_l$, w.r.t the probability distribution $p_l$ on $\{1,\cdots,\bar{N}\}$ (without replacement) and select $M_{\rm r} = M -M_{\rm t}$ indices uniformly at random (\ie random part). The probability distribution $p_l$ can take any arbitrary form, though in this paper we consider two candidates: \textit{(i)} TS w.r.t the intensity (TS-intensity) of the previous reconstructed layer $\hat{\bs x}_{l-1}$ and \textit{(ii)} TS w.r.t the gradient (TS-gradient) of the previous reconstructed layer. Mathematically,\
\begin{align}\label{eq:ts-pmf}
    {\rm (TS-intensity)} & &p_l(q) &= \frac{\hat{ x}_{l-1,q}}{\|\hat{\bs x}_{l-1}\|_1},\\
    {\rm (TS-gradient)} & &p_l(q) &= \frac{{\rm Grad}(\hat{\bs x}_{l-1})_q}{\|{\rm Grad}(\hat{\bs x}_{l-1})\|_1},
\end{align}
where $q\in \{1,\cdots,\bar{N}\}$ is the pixel index and the ${\rm Grad}(\cdot)$ operator takes a vector as an input, converts it into a matrix, computes and sums up the absolute value of horizontal and vertical gradients, and converts the outcome back into a vector. TS-intensity prioritizes bright locations while TS-gradient is in favour of sampling the edges. We note that the spatial distribution of the data cubes in FIB-SEM slowly varies from one layer to another; hence, the reconstructed image of a layer is a good proxy for the next layer. The same can also be applied to any scanning system where the features vary little between frames, such as STEM or SEM video, assuming the amount of drift between images is inconsequential.

The targeted part in our sampling strategy captures the static features, whereas the random part accounts for variations of the sample in the third dimension and mitigates imperfections in the recovery process. Moreover, as long as the reconstruction time of an image is less than the time that it takes for the FIB to remove the previous layer, such \textit{adaptive and targeted} sampling strategy is feasible in practice.

\subsection{Image recovery method}
We now focus on estimating $\hat{\bs x}_l \approx \bs x_l$ from subsampled measurements $\bs y_l^{\rm cs}$ in \eqref{eq:cs-fib-sem} for $l \in \{1,\cdots,N_3\}$, \ie the inpainting problem. In this work we assume that SEM images are sparse/compressible in an unknown dictionary that can be learned during the recovery process. This leads to the development of dictionary learning adopting a
Bayesian non-parametric method called Beta-Process Factor
Analysis (BPFA) as introduced in \cite{paisley2009nonparametric}. The advantages of this approach include the ability to infer both the noise variance and sparsity level of the signal in the dictionary, and allowing for the learning of dictionary elements directly from subsampled data. Since FIB-SEM acquisition is performed sequentially along the depth, we propose a layer-wise recovery of 2D images. 

Our recovery process adopted from \cite{sertoglu2015scalable} operates as follows. The FIB-SEM measurements $\bs y^{\rm cs}_l$ of layer $l \in \{1,\cdots,N_3\}$ are first partitioned into $N_p$ overlapping square patches $\{\bs y_{l,i}\}_{i=1}^{N_p}$ of size $B\times B$ pixels such that $\bs y_{l,i} \in \bb R^{B^2}$; resulting in $\ts N_p = (\sqrt{\bar{N}}-B+1)^2$ total number of patches. Similarly, we partition the corresponding sample surface image, mask operator, and noise as $\{\bs x_{l,i}\}_{i=1}^{N_p}$, $\{\bs P_{\Omega_{l,i}}\}_{i=1}^{N_p}$, and $\{\bs n^{\rm cs}_{l,i}\}_{i=1}^{N_p}$ respectively, such that for each patch $i \in \{1,\cdots,N_p\}$

\begin{equation} \label{eq:cs-fib-sem-patch-sensing-model}
    \bs y^{\rm cs}_{l,i} = \bs P_{\Omega_{l,i}} \bs x_{l,i} + \bs n^{\rm cs}_{l,i} \in \bb R^{B^2}.
\end{equation}

The patches of every surface image are assumed sparse in a shared dictionary, \ie $\bs x_{l,i} = \bs D_l \bs \alpha_{l,i}$, where $\bs D_l \in \bb R^{B^2\times K}$ denotes the dictionary with $K$ atoms and $\bs \alpha_{l,i} \in \bb R^{K}$ is a sparse vector of weights or coefficients for layer $l$. Note that we intend to learn a different dictionary for each layer. Based on these definitions, the BPFA algorithm allows us to infer $\bs D_l$, $\bs \alpha_l$, and the noise statistics and in turn reconstruct the layers of the sample in a sequential fashion.

BPFA operates based on the following assumptions. \textit{(i)} The dictionary atoms $\bs D_l \coloneqq \{\bs d_{l,k}\}_{k=1}^{K}$ are drawn from a zero-mean multivariate Gaussian distribution. \textit{(ii)} Both the components of the noise vectors $\bs n^{\rm cs}_{l,i}$ and the non-zero components of the weight vectors $\bs \alpha_l \coloneqq \{\bs \alpha_{l,i}\}_{i=1}^{N_p}$ are drawn \textit{i.i.d.} from zero-mean Gaussian distributions. \textit{(iii)} The sparsity prior on the weights is promoted by the Beta-Bernoulli process~\cite{paisley2009nonparametric}. Mathematically, by omitting the layer index $l$ for the sake of lighter notation, for all patches $i \in \{1,\cdots,N_p\}$ and dictionary atoms $k \in \{1,\cdots,K\}$,
\begin{subequations}
\begin{align}
    \bs y^{\rm cs}_i &= \bs P_{\Omega_i} \bs D \bs \alpha_i + \bs n^{\rm cs}_i, \!& \bs \alpha_i &= \bs z_i \circ \bs w_i \in \bb R^K,\label{eq:bpfa-1}\\
    \ts \bs D &= [\bs d_1^\top, \cdots, \bs d_{K}^\top]^{\top}, \!& \bs d_k & \sim  \cl N(0, B^{-2} \bs I_{B^2}),\label{eq:bpfa-2}\\
    \bs w_i & \sim  \cl N(0, \gamma_w^{-1} \bs I_{K}), \!& \bs n^{\rm cs}_i & \sim \cl N(0, \gamma_n^{-1} \bs I_{B^2}),\label{eq:bpfa-3}\\
    \bs z_i &\sim \!\ts \prod_{k=1}^{K} {\rm Bernoulli}(\pi_k),\ & \pi_k &\sim\! {\rm Beta}(\ts \frac{a}{K}, \!\ts\frac{b(K-1)}{K}),\label{eq:bpfa-4}
\end{align}
\end{subequations}
where $\bs I_K$ is the identity matrix of dimension $K$, $\circ$ denotes the Hadamard product, and $a$ and $b$ are the parameters of the Beta-process. The binary vector $\bs z_i$ in \eqref{eq:bpfa-4} determines which dictionary atoms to be used to represent $\bs y_i$ or $\bs x_i$; and $\pi_k$ is the probability of using a dictionary atom $\bs d_k$. In \eqref{eq:bpfa-3}, $\gamma_w$ and $\gamma_n$ are the (to-be-inferred) precision or inverse variance parameters. The sparsity level of the weight vectors, $\{\|\bs \alpha_i\|_0\}_{i=1}^{N_p}$ is controlled by the parameters $a$ and $b$ in \eqref{eq:bpfa-4}. However, as discussed in \cite{zhou2009non}, those parameters tend to be non-informative and the sparsity level of the weight vectors is inferred by the data itself.

Unknown parameters in the model above are inferred using stochastic Expectation Maximisation (EM) \cite{dempster1977maximum,sertoglu2015scalable}. In short, EM involves an expectation step to form an estimation of the latent variables, \ie $\{\|\bs \alpha_i\|_0\}_{i=1}^{N_p}$, and a maximisation step to perform a maximum likelihood estimation to update other parameters.
\section{Numerical results}\label{sec:simulations}

\begin{figure}[t]
    \centering
    \scalebox{1.2}{\includegraphics{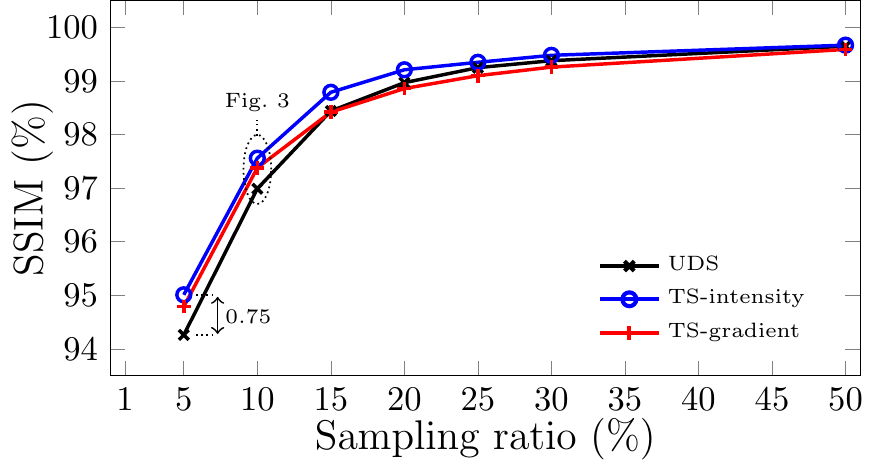}}
    \caption{Performance comparison between the two proposed sampling strategies and UDS over a range of sampling ratios.} 
    \label{fig:ssim_samplingratio}
\end{figure}

\begin{figure*}[t!]
    \centering
    \scalebox{0.83}{\includegraphics{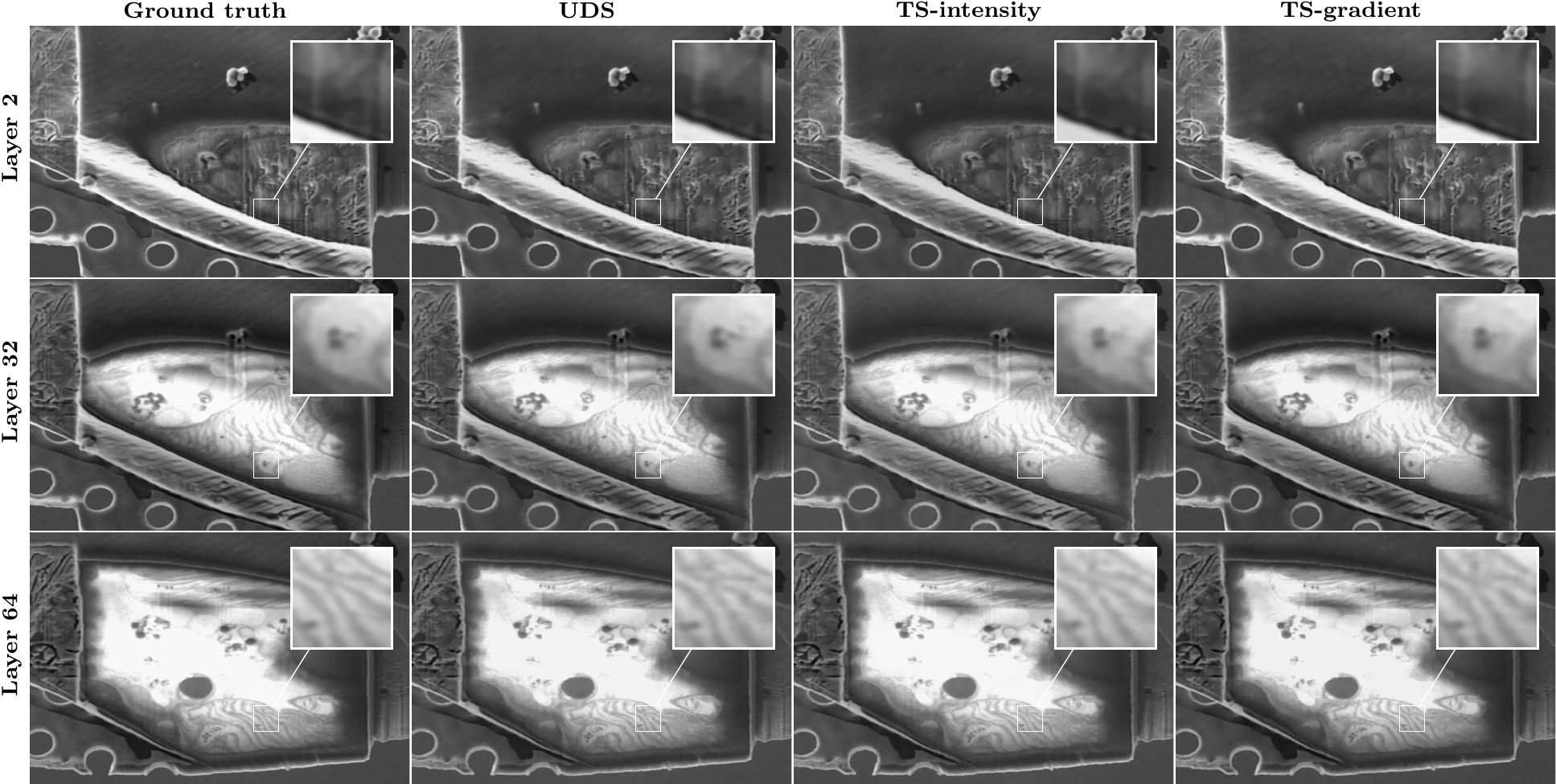}}
    \caption{Examples of treated data from \textit{E. gracilis} data cube. The ground truth and corresponding reconstructed images from 10\% of the measurements are shown. Visually, all three methods of subsampling provide good quality reconstructions. Targeted sampling provides slightly increased sharpness and spatial fidelity.}
    \label{fig:example}
\end{figure*}

\noindent \textbf{Experimental setup.} The data cube used in this paper is a cryo FIB-SEM tomography data set of a \textit{Euglena gracilis} cell acquired using using a JEOL JIB-4700F Z FIB-SEM (JEOL, Japan), equipped with a Leica micro-systems VCT500 cryo stage and cryo transfer system (Leica micro-systems, Austria). The acquisition took 12.4 hours in total: approximately three minutes for each cut and 37 seconds for each image acquisition.

\vspace{1mm}
\noindent\textbf{Simulation setup.} A set of 64 consecutive frames were selected from the larger \textit{E. gracilis} data set described above; resulting in a data cube of size $(N_1,N_2,N_3) = (1280,960,64)$. This FIB-SEM data was then denoised using BPFA and the resulting data cube was considered as ground-truth. The artificial compressive cryo FIB-SEM measurements were generated using the UDS and two proposed sampling strategies and were inpainted layer-by-layer using BPFA. This process was repeated over five independent realisations of the subsampling masks. For all of the data, the mean reconstruction time was 7.3 $\pm$ 0.6 seconds. The BPFA parameters used were $K = 36$, $B=14$, $N_{\rm epoch} = 2$, $\eta = 0.85$, $N_{\rm batch} = 163844$, $a = 1$, $b = 0$, $\gamma_n = 1$, $\gamma_{\omega}=10^6$, where $N_{\rm epoch}$ is the number of passes over all patches, $\eta$ is the learning rate and $N_{\rm batch}$ is the batch size used for each mini-batch operation.

\vspace{1mm}
\noindent\textbf{Results.} 
Fig.~\ref{fig:ssim_samplingratio} shows the average Structural Similarity Index Measure (SSIM) of all 64 layers as a function of sampling percentage. This shows TS-intensity as the leading method, followed by UDS at sampling ratios above 15\% and TS-gradient below 15\%. Interestingly, this method performs best at lower sampling ratios, meaning this method is particularly useful for samples which are sensitive to beam damage, such as biological specimens. Intensity-based targeted sampling provided up to a 0.75 increase in SSIM and gradient-based targeted sampling provided up to a 0.53 increase in SSIM at no operational cost when compared to UDS sampling. 

As the cut time for each FIB cross-section was three minutes for this sample, this current implementation is viable as a technique to operate without interruption. Operating at 10\% sampling, where the incident electron dose rate would also be a tenth of the full acquisition, the total acquisition time for this data set would theoretically be 10.5 hours, a 16\% reduction compared to actual the 12.4 hour acquisition time. All reconstructions were performed under Windows 11 (x64) with an Intel Core i5-6500 CPU and NVIDIA GeForce RTX 3090 GPU. 

Fig.~\ref{fig:example} shows various reconstructions from the reconstruction 10\% data set. The 2nd, 32nd, and 64th layers are shown to help visualise the data cube and show the method applied to sections of the data cube with different structures present. For each layer shown, the ground truth, UDS, TS-intensity, and TS-gradient reconstructions are presented. The \textit{E. gracilis} cell is the bright object at the centre of each frame, with the surrounding detail being the support structure.

\section{Conclusion}
\label{sec:conclusion}

This proof of concept work shows compressive sensing with targeted sampling to be a viable method for increasing reconstruction quality for a subsampled cryo FIB-SEM data set. This novel approach will help overcome limitations in data acquisition with regards to beam sensitive samples if implemented experimentally. While an improvement is seen when using targeted sampling, UDS still provides value as a sampling method. 

It is posited that data sets with higher frequency structures or structures with higher contrast may see increased effectiveness of TS compared to the \textit{E. gracilis} data set, in which individual images are relatively smooth. 

Future work aims to test this method applied to other data sets, as well as experimental implementation to validate the method's efficacy in practice. This future study would also aim to address the assumptions made here with regards to beam damage and scan coil deflector hysteresis, as well as test other potential sampling strategies.

\vfill\pagebreak 

\bibliographystyle{IEEEtran}
\bibliography{CSCRYOSEM}

\end{document}

%% file: packages.tex
\usepackage{amssymb,amsthm,amsmath,array}
\usepackage[top=1in, left=1in, right=1in, bottom=1in]{geometry}
\usepackage{graphicx}
\usepackage[caption=false,font=footnotesize]{subfig}
\usepackage{xspace}
\usepackage[sort&compress, numbers]{natbib}
\usepackage{stmaryrd}
\usepackage{xcolor}
\usepackage{todonotes}
\usepackage{mathtools}
\usepackage{float}
\usepackage{textcomp}

\graphicspath{{Figures/}}

\usepackage{tikz}
\usetikzlibrary{external}
\usetikzlibrary{arrows}
\usepackage{pgfplots}
\pgfplotsset{compat=1.7}

\newcommand{\bs}{\boldsymbol}
\newcommand{\bb}{\mathbb}
\newcommand{\cl}{\mathcal}

\newcommand{\ts}{\textstyle}
\newcommand{\ie}{\emph{i.e.},\xspace}
\newcommand{\eg}{\emph{e.g.},\xspace}

\renewenvironment{abstract}{\bf {\em\ Abstract---}}{}

\makeatletter
\renewcommand{\section}{\@startsection {section}{1}{\z@}%
             {-3.5ex \@plus -1ex \@minus -.2ex}%
             {2.3ex \@plus.2ex}%
             {\normalfont\large\bfseries}}
\makeatother